\documentclass[12pt]{article}
\usepackage{amsmath}
\usepackage{amssymb}
\usepackage{amsfonts}
\newcommand{\be}{\begin{equation}}
\newcommand{\ee}{\end{equation}}
\newcommand{\bea}{\begin{eqnarray}}
\newcommand{\eea}{\end{eqnarray}}
\newcommand{\ep} {\epsilon}

\usepackage{color}

\usepackage{epsfig}
\usepackage{graphicx}
\usepackage{hyperref}
\newcommand{\email}[1]{\footnote{#1}}
\begin{document}
\baselineskip=22pt
%
%PREPRINT NUMBERS
\begin{flushright}
\hfill{\bf USTC-ICTS-04-04}\\
\end{flushright}
%
%
%TITLE AND AUTHORS
\begin{center}
{\large \bf Maximally symmetric subspace decomposition of the Schwarzschild black hole}
\vspace*{1cm}

Shu Dai\email{sprinna@mail.ustc.edu.cn} and Cheng-Bo Guan\email{guancb@ustc.edu.cn}\\
{\em Interdisciplinary Center for Theoretical Study}\\
{\em University of Science and Technology of China, Hefei, Anhui 230026, China}
\end{center}
\vspace*{1.0cm}

%
%ABSTRACT
\begin{abstract}
The well-known Schwarzschild black hole was first obtained as a stationary,
spherically symmetric solution of the Einstein's vacuum field equations.
But until thirty years later, efforts were made for the analytic extension
from the exterior area $(r>2GM)$ to the interior one $(r<2GM)$. As a contrast
to its maximally extension in the Kruskal coordinates, we provide a comoving
coordinate system from the view of the observers freely falling into the
black hole in the radial direction. We find an interesting fact that the
spatial part in this coordinate system is maximally symmetric $(E_3)$, i.e.,
along the world lines of these observers, the Schwarzschild black hole can be
decomposed into a family of maximally symmetric subspaces.
\end{abstract}
\begin{flushleft}
\end{flushleft}
\newpage
%
%
%Introduction
\section{Introduction}
Shortly after Einstein published {\it The Foundation of the General Theory of Relativity},
the description of the exterior gravitational field around a static, spherically symmetric body
was found by Schwarzschild\cite{Schwarz}, as a solution of the Einstein's vacuum field equations,
\be\label{stand}
ds^2=(1-r^*/r)dt^2-(1-r^*/r)^{-1}dr^2-r^2(d\theta^2+\sin^2{\theta}d\varphi^2)~,
\ee
where $r^*=2GM$~.
Discussions on the exterior solution $(r>r^*)$ helped greatly to the tests of the general
relativity, such as the gravitational redshift, the bending of light and the precession
of Mercury's orbit, etc. As we have known, in the above standard coordinate system (from the
view of the observers at infinity), the metric singularity at the Schwarzschild radius
$(r=r^*)$ is not a real space-time singularity. This means an object can freely
pass through the null spherical surface (with the Schwarzschild radius) into the interior
area. For the ordinary celestial bodies like the Sun, the Schwarzschild radius lies
deep in the interior of these bodies, and only outside can the vacuum Schwarzschild
solution be applicable. However, for a super massive celestial body, particularly
when its nuclear fuel has been burnt out, it will inevitably collapse till a trapped
surface (like that of the Schwarzschild solution) forms. And then comes into
being a black hole\cite{BHS}-\cite{Joshi}.

To study the global (causal) structure of the space-time is of great importance in the general
relativity\cite{HWK,WLD}. As a coordinate system maybe only cover one region of the
space-time, it is quite necessary to analyse the possible extension from this region,
even though there exist metric singularities at the boundary. In the theory of general
relativity, since there is no preclusion for the creation of black holes, it should not be
meaningless to study the physical behaviors of the interior area. The analytic extension from
the exterior to the interior is therefore necessary. The maximal extension of the
Schwarzschild solution was finally provided by Kruskal\cite{Krus} in a conformally flat
(null) coordinate system,
\bea
&&ds^2=f^2(r)(dv^2-du^2)-r^2(d\theta^2+\sin^2{\theta}d\varphi^2)~,\\
&&f^2(r)=(4r^{*3}/r)\exp(-r/r^*)~,\nonumber\\
&&u(t,r)=(r/r^*-1)^{1/2}\exp(r/2r^*)\cosh(t/2r^*)~,\nonumber\\
&&v(t,r)=(r/r^*-1)^{1/2}\exp(r/2r^*)\sinh(t/2r^*)~.\nonumber
\eea
This coordinate system covers analytically the whole Schwarzschild black hole
except the space-time singularity at the origin $(r=0)$.

Nevertheless, the time coordinates of the above systems are not proper time.
With this intension, we want to seek a coordinate system of the Gaussian normal form,
\be
ds^2=d\tau^2-h^2(\tau,\rho)d\rho^2-r^2(d\theta^2+\sin^2{\theta}d\varphi^2)~.
\ee
It is easy to see that the time coordinate variable $\tau$ is just the proper time,
which can be viewed as the readings of clocks freely falling into the black hole along
the radial direction. By solving the coordinate transformation and the function
$h(\tau,\rho)$, we find that this coordinate system is also an analytic extension
of the exterior Schwarzschild solution. Furthermore, given a fixed time $\tau$, the spatial section
is just the three-dimensional Euclidean space $E_3$. That is to say, we have a
maximally symmetric subspace decomposition for the Schwarzschild black hole.

This paper is organized as follows. In Section 2, the discussion for the radial motion
of a freely falling particle is briefly reviewed. Making use of its results, we provide
the explicit coordinate transformation and the function $h(\tau,\rho)$ in Section 3.
Moreover, some interesting facts about this new comoving coordinate system are
demonstrated. In the end, our conclusions are given.
\section{Radial motion of a freely falling particle}
Neglecting its self-gravitational field, we know the world line of a freely
falling particle is generally a geodesic determined by the background space-time,
\be
\frac{d^2x^{\mu}}{dp^2}-
\Gamma^{\mu}_{\lambda\sigma}\frac{dx^{\lambda}}{dp}\frac{dx^{\sigma}}{dp}=0~.
\ee
For the exterior Schwarzschild solution, we now investigate the freely radial motion in
the standard coordinate system $(\ref{stand})$,
\be
\begin{array}{l}
\displaystyle\frac{d\theta}{dp}=0~,~~\frac{d\varphi}{dp}=0~,\\[0.4cm]
\displaystyle\frac{d^2t}{dp^2}+\frac{A^{\prime}}{A}\cdot\frac{dt}{dp}\frac{dr}{dp}=0~,\\[0.4cm]
\displaystyle\frac{d^2r}{dp^2}+\frac{1}{2}AA^{\prime}\cdot\left(\frac{dt}{dp}\right)^2-
\displaystyle\frac{1}{2}\frac{A^{\prime}}{A}\cdot\left(\frac{dr}{dp}\right)^2=0~,
\end{array}
\ee
where $A(r)=1-r^*/r$ and $A^{\prime}$ denotes the derivative with respect to $r$. By the
first integral to the last two equations, we have conversed integral constants as,
\be
A\cdot\frac{dt}{dp}=\ep~,
\ee
and
\be\label{kappa}
\frac{1}{A}\cdot\ep^2-\frac{1}{A}\cdot\left(\frac{dr}{dp}\right)^2=\kappa~.
%A\cdot\left(\frac{dt}{dp}\right)^2-\frac{1}{A}\cdot\left(\frac{dr}{dp}\right)^2=\kappa~.
\ee
Considering a time-like geodesic, and choosing the proper time $\tau$ as the world line
parameter $p$, we will have a normalized velocity vector,
\be
1=U^{\mu}U_{\mu}=g_{\mu\nu}\frac{dx^{\mu}}{d\tau}\frac{dx^{\nu}}{d\tau}=
A\cdot\left(\frac{dt}{d\tau}\right)^2-\frac{1}{A}\cdot\left(\frac{dr}{d\tau}\right)^2=\kappa~.
\ee
This means that, for the radial time-like geodesic motion, the integral constant $\kappa$ can
be normalized by reparameterization.

In fact, the integral constant $\ep$ can be considered to be the particle's energy
that is kept invariant along the geodesic $(d\ep/d\tau=0)$. Besides, taking the limit
of $``r\rightarrow \infty"$ for the equation $(\ref{kappa})$, one will have,
\be
\ep^2\cdot\left[1-\left({dr}/{dt}\right)^2|_{r\rightarrow\infty}\right]=
\ep^2\cdot(1-v_0^2)=1~.
\ee
This indicates that the conserved energy $\ep$ is just relevant to the initial
velocity of the falling particle at infinity. And therefore, for an initially
resting particle, one has $\ep=1$. This will be our foundation to set up the
Gaussian normal coordinate system in the next section.
\section{Gaussian normal coordinate system}
Considering the coordinates comoving with the circumferentially distributed observers,
who are asymptotically rest at infinity and freely fall into the Schwarzschild black
hole in the radial direction, we can suppose the metric in this coordinate system is
of the following form,
\be\label{Gauss}
ds^2=d\tau^2-h^2(\tau,\rho)d\rho^2-r^2(d\theta^2+\sin^2{\theta}d\varphi^2)~.
\ee
At the same time, from the analysis in the above section, we immediately have
the differential relations between the coordinate variables $(t,r)$ and the
proper time $\tau$,
\be
\begin{array}{l}
\displaystyle\frac{\partial{t}}{\partial\tau}=(1-r^*/r)^{-1}~,\\[0.4cm]
\displaystyle\frac{\partial{r}}{\partial\tau}=-\sqrt{r^*/r}~.
\end{array}
\ee
Inserting this two relations into equation $(\ref{stand})$, and noticing the equivalence
of the two line elements $(\ref{stand})$ and $(\ref{Gauss})$, we will obtain
the differential relations between the coordinate variables $(t,r)$ and
the spatial coordinate variable $\rho$,
\be
\begin{array}{l}
\displaystyle\left(\frac{\partial{t}}{\partial\rho}\right)^2=
             \frac{r^*}{r}\cdot\frac{h^2}{(1-r^*/r)^2}~,\\[0.4cm]
\displaystyle\left(\frac{\partial{r}}{\partial\rho}\right)^2=h^2~.
\end{array}
\ee
In addition to these constraints, there should exist the integrable conditions for this
coordinate transformation,
\be
\begin{array}{l}
\displaystyle\frac{\partial^2{t}}{\partial\tau\partial\rho}=
             \frac{\partial^2{t}}{\partial\rho\partial\tau}~,\\[0.4cm]
\displaystyle\frac{\partial^2{r}}{\partial\tau\partial\rho}=
             \frac{\partial^2{r}}{\partial\rho\partial\tau}~.
\end{array}
\ee
As a result, one gets the constraint on the function $h$,
\be
\frac{\partial{h}}{\partial\tau}=\frac{1}{2}h\sqrt{r^*/r^3}~.\nonumber
\ee
One simple solution is,
\be
h=\sqrt{r^*/r}~~{\rm and}~~h^2=r^*/r~.
\ee
Now the coordinate transformations can also be easily given,
\be\label{Trans}
\begin{array}{l}
\displaystyle~r^{3/2}=-\frac{3}{2}r^{*1/2}(\tau+\rho)~,\\[0.4cm]
\displaystyle~t=\tau+r^*\left(\ln{\frac{\sqrt{r/r^*}+1}{\sqrt{r/r^*}-1}}
                    -2\sqrt{r/r^*}\right)~.
%\frac{1}{2}(\tau-\rho)+r^*\left(\ln{\frac{\sqrt{r/r^*}+1}{\sqrt{r/r^*}-1}}-
%                \frac{1}{3}(r/r^*)^{3/2}-2\sqrt{r/r^*}\right)~.
\end{array}
\ee
At last, we obtain the metric of the Gaussian normal form,
\be
ds^2=d\tau^2-\frac{r^*}{r}d\rho^2-r^2(d\theta^2+\sin^2{\theta}d\varphi^2)~.
\ee
As we can see, there is no singularity in this metric except those points at $r=0$.
Although the above discussion is only on the exterior region, it is not difficult to
see that, this comoving coordinate system can extend smoothly to the interior region.
And those broken geodesics in the standard coordinate system can also go freely through
the null surface into the interior area. In figure 1, we give an illustration for
this coordinate system, and show its nontrivial extension.

%***********************************************************
%FIGURE OF DEMONSTRATION FOR THE GAUSSIAN NORMAL COORDINATES
\begin{figure}[htb]\label{pic}
\begin{center}
\includegraphics[height=80mm,width=80mm]{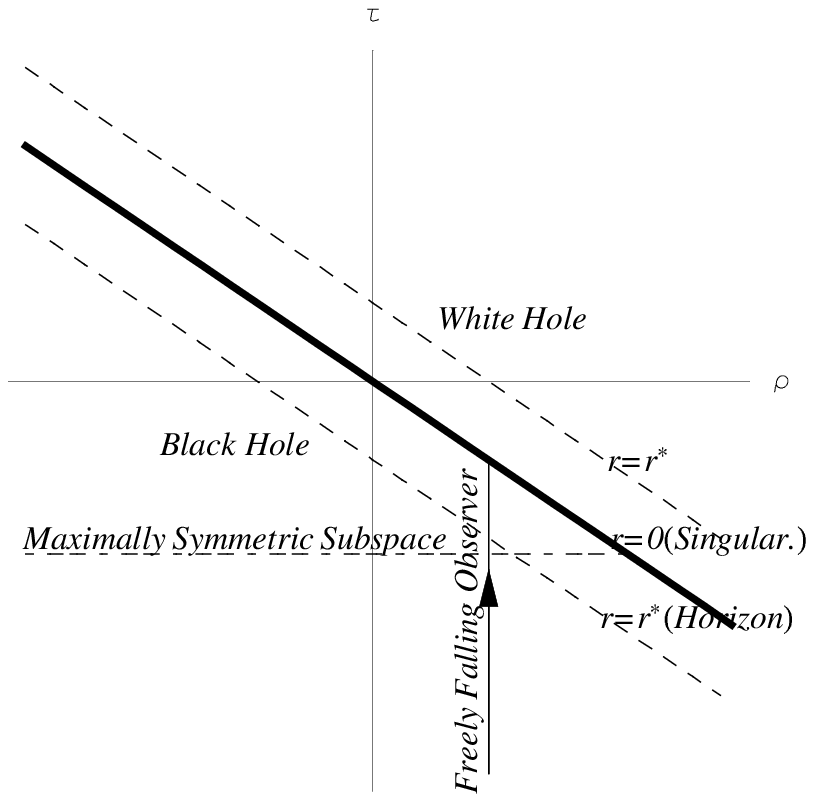}
\caption{\scriptsize The Gaussian normal coordinate system is also an analytic (maximal)
                     extension of the Schwarzschild black hole. As shown in this sketch,
                     the lower-left region $(\tau+\rho\leq 0)$ describes a black hole, and
                     the upper-right region $(\tau+\rho\geq 0)$ represents a white hole.
                     Fixing one point of time, the spatial section is maximally symmetric
                     except the singularity at $r=0$.}
\end{center}
\end{figure}
%
%***********************************************************

Now let's take a certain proper time $\tau=\tau_0$. The spatial section at this moment
can be characterized by this line element,
\be\label{Euc1}
d\bar{s}^2=\frac{r^*}{r_0}d\rho^2+r_0^2(d\theta^2+\sin^2{\theta}d\varphi^2)~,~~~
r_0=r(\tau_0,\rho)~.
\ee
In view of the transformation relation $(\ref{Trans})$, one has
\be
r_0dr_0^2=r^*d\rho^2~, \nonumber
\ee
which then reduces the above line element to a familiar form,
\be
d\bar{s}^2=dr_0^2+r_0^2(d\theta^2+\sin^2{\theta}d\varphi^2)~.
\ee
This is well-known as the metric of three-dimensional Euclidean space $(E_3)$.
Therefore, along the proper time of the freely falling observers, the Schwarzschild
black hole can be decomposed into subspaces of maximal symmetry.  It is should be
pointed out however, that there still exits a singularity at the origin $(r_0=0)$,
which is shown clearly in equation $(\ref{Euc1})$ and demonstrated in figure 1.
%
%In fact, as can be seen from  the spatial section do
%
%\newline
\section{Conclusions}
Taking the coordinate time as the proper time carried by the freely falling observers,
we set up a Gaussian normal coordinate system. This comoving coordinate system is also
an analytic extension from the exterior region to the interior one. We also find
that the spatial part with a fixed proper time is the well-known Euclidean space (except
a singularity). It is therefore concluded that, at least locally, the Schwarzschild
black hole can be decomposed into maximally symmetric subspaces along the proper coordinate time.
\newline
\newline
%Acknowledgement
%\newpage
\noindent
{\large\bf Acknowledgement:}\newline
This work is supported partly by the Natural Science Foundation of
China. One of us (C.B. Guan) is also supported by grants through the
ICTS (USTC) from the Chinese Academy of Sciences.
%
%
%References

%
%
%THE END
\end{document}